\documentclass[12pt]{article}
\usepackage{amsfonts,graphicx}
\newcommand{\beq}{\begin{equation}}
\newcommand{\eeq}{\end{equation}}
\newcommand{\beqa}{\begin{eqnarray}}
\newcommand{\eeqa}{\end{eqnarray}}
\newcommand{\beqar}{\begin{eqnarray*}}
\newcommand{\eeqar}{\end{eqnarray*}}

\newcommand{\eg}{{\it e.g.,}\ }
\newcommand{\ie}{{\it i.e.,}\ }

\newcommand{\norm}[1]{\raise.3ex\hbox{:}#1\raise.3ex\hbox{:}}

\newcommand{\labell}[1]{\label{#1}}
\newcommand{\reef}[1]{(\ref{#1})}
\newcommand{\Tr}{{\rm Tr}}
\newcommand{\STr}{{\rm STr}}

\newcommand\hi{{\rm i}}

\newcommand\prt{\partial}

\newcommand\ls{\ell_s}

\newcommand\hR{{\hat{R}}}

\newcommand\tF{{\tilde F}}
\newcommand\tB{{\tilde B}}
\newcommand\tG{{\tilde G}}
\newcommand\dtF{{{^*\!\tilde F}}}
\newcommand\vB{\vec{B}}
\newcommand\nab{\vec{\nabla}}
\newcommand\identity{{\mathbf{1}}}
\newcommand\bbR{{\mathbb{R}}}

\begin{document}

\thispagestyle{empty}
\rightline{\small hep-th/0102080 \hfill McGill/01-01}
\vspace*{2cm}

\begin{center}
{\bf \LARGE Non-abelian Brane Intersections}
\vspace*{1cm}

Neil R. Constable\footnote{E-mail: constabl@hep.physics.mcgill.ca},
Robert C. Myers\footnote{E-mail: rcm@hep.physics.mcgill.ca}
and \O yvind Tafjord\footnote{E-mail: tafjord@physics.mcgill.ca}

{\it Department of Physics, McGill University}\\
{\it Montr\'eal, QC, H3A 2T8, Canada}\\
\vspace{2cm} ABSTRACT
\end{center}
We study new solutions of the low-energy equations of motion
for the non-abelian D-string. We find a ``fuzzy funnel'' solution
consisting of a noncommutative four-sphere geometry which expands
along the length of the D-string. We show
that this funnel solution has an interpretation as D-strings
ending on a set of orthogonal D5-branes. Although not supersymmetric,
the system
appears to be stable within this framework. We also give a dual
description of this configuration as a bion spike in
the non-abelian world volume theory of coincident D5-branes.

\vfill \setcounter{page}{0} \setcounter{footnote}{0}
\newpage

\section{Introduction}

The advent of D-branes~\cite{joe} has brought about significant advances in
string theory --- for reviews see refs.~\cite{tasi,clifford}. Much of the
progress has come about by directly studying the low energy dynamics 
of the D-brane's world volume which is known to be governed by the 
Born-Infeld action~\cite{bin,yet2}. This world volume theory living 
on D-branes has many fascinating
features. Among these is the possibility for D$p$-branes, through an
appropriate excitation of fields, to morph into objects resembling
D$p'$-branes of lower or higher dimensionalities. This is seen, for
instance, in the Born-Infeld string \cite{calmaldetc} where a D3-brane, pulled
out into a spike and adorned with a magnetic monopole field, describes
a D-string extending out from the D3-brane. We also know from matrix theory
\cite{matrix,wati2} and the dielectric effect for D-branes \cite{dielec}
that a collection of D0-branes can expand into a D2-brane.
This is a direct consequence of the fact that the world
volume theory living on a collection of coincident D-branes possesses a 
non-abelian gauge symmetry~\cite{bound} and that the world volume fields 
describing the brane's position become matrix valued in this context. 
The non-abelian generalization of the 
Born-Infeld action has received much attention recently.
The form of the bosonic part of the
 action for arbitrary supergravity backgrounds
was proposed
in refs.~\cite{dielec,watiprep} 
and has led to many interesting developments including
the giant gravitons of ref.~\cite{giant}, the resolution of curvature
singularities {\it a la} Polchinski and Strassler~\cite{PS} as well as several
new insights into
world volume gauge theories as described, for example, in 
refs.~\cite{bioncore,lam1,zam,coset,pioline,kluson,lozano}. 

In this paper, we will use the non-abelian Born-Infeld action to
study a set of D1-branes ending on a collection of
orthogonal D5-branes. We find that this D5$\perp$D1 intersection shares many
features with the well-known D3$\perp$D1 system, but there will, of course, be
interesting  differences as well. Since the D5$\perp$D1-system is not
a brane intersection that is commonly discussed, let us begin by reminding
the reader how it is possible for D-strings to end on D5-branes.
The D-strings are electric charge sources for the Ramond-Ramond (RR) two-form
potential $C^{(2)}$, and so one must understand how this charge is conserved
in the D5$\perp$D1 intersection \cite{strominger}. In the presence of
D5-branes, the equations of motion for the RR two-form are modified by
the appearance of new source terms due to the following term
in the D5-brane action,
\beq
S_{WZ}\simeq 2\pi^2\alpha'^2\mu_5\int_{{\rm D}5}\STr\left[ C^{(2)}\wedge
F\wedge F\right].
\labell{swz}
\eeq
Hence D-strings can end on D5-branes, but they must act as sources of
second Chern class or instanton number  in the world volume theory
of the D5-branes \cite{semenoff}. That is, the intersection will be
described by a configuration involving the D5-brane gauge field $F$ with 
$\int F\wedge F \neq 0$, 
for any four-surface surrounding the D-string endpoints. Hence, one interesting
point that distinguishes this from the D3$\perp$D1 system, is that if
the D5$\perp$D1 intersection is to be resolved by a smooth low
energy field configuration, it must involve a {\it non-abelian} gauge
field. This, in turn, indicates that the intersection must involve several
coincident D5-branes rather than just a single D5-brane. Another feature,
which distinguishes the D5$\perp$D1 intersection from the D3$\perp$D1 system,
is that it is {\it not} supersymmetric. The latter is easily seen in
the context of perturbative string theory. The D5-D1 strings would have
six Neumann-Dirichlet directions, which is inconsistent with a supersymmetric
ground-state \cite{tasi,clifford}.

In analogy with the previous work on the D3$\perp$D1 case
\cite{calmaldetc,ded, bioncore}, we will examine the D5$\perp$D1 configuration
from two dual points of view. Beginning with the world volume theory 
of the D-strings, in analogy to the D3$\perp$D1 construction in 
ref.~\cite{bioncore}, we find a funnel configuration in which the 
D-strings expand through a noncommutative geometry into a set of D5-branes.
The second point of view is constructing the
D5$\perp$D1 intersection within the D5-brane world volume theory.
Here we construct a non-abelian gauge field as described above, and we also
excite one of the transverse scalars. The D-strings then appear as
a bion spike which smoothly deforms the (otherwise) planar geometry
of the D5-branes. 

The noncommutative space which we encounter in the D-string funnel
is the fuzzy four sphere, which we
describe using an explicit matrix 
representation found in ref.~\cite{wati4}. 
Due to a curious
restriction on the dimensions of these matrices we find that the number
of D-strings, $N$, and the number of D5-branes, $n$, are related by 
$N\sim n^3/6$ for large $n$. 
We find a similar relation in the D5-brane world volume description. There 
our construction of the D5$\perp$D1 intersection 
involves $SU(n)$ instantons which are homogeneous on the four sphere. 
It turns out that the 
spherical symmetry imposes an upper bound on the instanton number, $N$,
given again by $N \sim n^3/6$ for large $n$! 

We find also that quantities such as the energy and Ramond-Ramond charges
of the D5$\perp$D1 intersection agree between the dual D-string and 
D5-brane points of view at large $N$. While the agreement begins to fail for 
small $N$ this should not be
viewed as a failure of the duality between these two descriptions. Rather
it is a limitation of our analysis which relies on the non-abelian Born-Infeld
action. The latter is only a {\it low energy} action which does not give
a reliable description of all field configurations. Hence, for small $N$,
our dual analyses give complimentary descriptions of the same system which are
valid in different regimes. Of course, similar limitations are also found
in examining the D3$\perp$D1 intersection \cite{bioncore}.

An outline of the paper is as follows: Since many of our
considerations are in analogy with previous studies of the
D3$\perp$D1 system, we give a brief review of the latter in Section~2.
In Section~3, we consider the D5$\perp$D1 system from the
point of view of the low energy D-string theory, while in Section~4, we do the
same using the D5-brane world volume theory. In Section~5, we present the
extension of our analysis to the case where the D-strings are replaced by
dyonic strings, \ie bound states of fundamental and Dirichlet strings.
In Section 6, we consider fluctuations of the static D5$\perp$D1 configuration
from the point of view of the D-string theory. Finally in
Section~7, we provide a discussion of our results. Appendix A
summarizes the properties of the matrices used to construct the
noncommutative four-sphere.
In Appendix B, we give a discussion of the homogeneous instantons which
play a role in the D5-brane construction of Section~4.

\section{Review of the D3$\perp$D1-brane system}

In this section, we briefly review a system consisting of a set of
D-strings ending on an orthogonal D3-brane. This has two dual descriptions
coming from the D3-brane and the D-string world volume theories. 
Within either of these theories, the field configurations can be determined
by simply solving the BPS conditions or by studying the full 
Born-Infeld equations of motion. Here, we use a simple approach based
on minimizing the energy \cite{gauntlett},
while pointing out some aspects that will be of relevance in later
sections.

We start with the D3-brane theory construction.
The low energy dynamics of 
a (single) D3-brane in a Minkowski space background is described
by the Born-Infeld action \cite{bin},
\beq
S=\int dt\, {\cal L}=-T_3\int d^4\sigma\sqrt{-\det\left(\eta_{ab}+
\lambda^2\partial_a\phi^i\partial_b\phi^i+\lambda F_{ab}\right)},
\labell{act3}
\eeq
where $\lambda\equiv2\pi\alpha'=2\pi\ls^2$. Here
we have implicitly used static gauge, with $\sigma^a\,(a=0,\ldots,3)$ denoting
the world volume coordinates while $\phi^i\,(i=4,\ldots,9)$ are the
scalars describing transverse fluctuations of the brane. Also, $F_{ab}$ is the
field strength of the $U(1)$ gauge field on the brane.

Within this theory, D-strings appear as BPS magnetic monopoles
\cite{calmaldetc}. Hence we consider exciting one of the scalar
fields, say $\phi\equiv\phi^9$, as well as a magnetic field
$B^r={1\over2}\epsilon^{rst}F_{st}\ (r,s,t=1,\ldots,3)$. One can show
that it is consistent with the equations of motion to
set all the other fields equal to zero. For static configurations, we then
evaluate the energy,
\beqa
E&=&-{\cal L}=T_3\int d^3\sigma\sqrt{1+\lambda^2|\nab\phi|^2
+\lambda^2|\vB|^2+\lambda^4(\vB\cdot\nab\phi)^2}\nonumber\\
&=&T_3\int d^3\sigma\sqrt{\lambda^2|\nab\phi\mp\vB|^2+
(1\pm\lambda^2\vB\cdot\nab\phi)^2}\nonumber\\
&\ge&T_3\int d^3\sigma\left(1\pm\lambda^2\vB\cdot\nab\phi\right).
\label{enineq}
\eeqa
The first term in this lower bound is simply the energy of the D3-brane.
The second term
is a total derivative: $\vB\cdot\nab\phi=\nab\cdot(\vB\phi)$ using
the Bianchi identity $\nab\cdot\vB=0$. Hence this term is topological,
depending only on the boundary values of the fields specified at infinity
and near singular points (after introducing a cutoff, such as at $r=0$ below).
Therefore, the last line of eq.~(\ref{enineq}) provides a true minimum 
of the energy for a given set of boundary conditions. 

The lower bound in eq.~\reef{enineq} is achieved when
\beq
\nab\phi=\pm\vB\ ,
\labell{ssusy}
\eeq
which coincides with the BPS condition for the magnetic monopoles
\cite{calmaldetc}. Further, using the Bianchi identity, eq.~\reef{ssusy}
implies $\nab^2\phi=0$ and so we are looking for harmonic functions
on the D3-brane. The simplest solution, corresponding to the ``bion spike,''
is given by
\beq
\phi(r)={N\over 2r},\qquad\quad
\vB(\vec{r})=\mp{N\over 2r^3}\vec{r},
\label{phisol}
\eeq
where $r^2=(\sigma^1)^2+(\sigma^2)^2+(\sigma^3)^2$, and
$N$ is an integer due to quantization of the magnetic charge. The
energy of this configuration is readily computed to be
\beq\labell{d3d1energy}
E=T_3\int d^3\sigma+N T_1\int_0^\infty d(\lambda\phi),
\eeq
where $T_1=(2\pi \ls)^2 T_3$.
Hence we have recovered the energy expected for a BPS
configuration consisting of $N$ semi-infinite D-strings ending
on an orthogonal D3-brane. Note that the physical distance in the transverse
direction is $\lambda\phi$.
Other considerations, such as the effective charge
distribution and fluctuations on the strings \cite{waves}, confirm that
the bion spike describes the D3$\perp$D1-system to a good approximation.

Now we consider the dual description of the D3$\perp$D1-system
given by the D-string theory.
The low energy dynamics of $N$
D-strings in a flat background is well described by the
non-abelian Born-Infeld action \cite{dielec,yet2}:
\beq
S=-T_1\int d^2\sigma\, \STr\sqrt{-\det\left(\eta_{ab}+
\lambda^2\partial_a\Phi^i Q^{-1}_{ij}\partial_b\Phi^j\right)
\ \det\left(Q^{ij}\right)}\ ,\labell{action}
\eeq
where $Q^{ij}=\delta^{ij}+i\lambda[\Phi^i,\Phi^j].$
Again we are assuming static gauge where the two worldsheet coordinates are
identified with $\tau=x^0$ and $\sigma=x^9$. The transverse scalars
$\Phi^i\ (i=1,\ldots,8)$ are now $N\times N$ matrices transforming
in the adjoint representation of the $U(N)$ worldsheet gauge symmetry.
The symmetrized trace prescription~\cite{yet}, denoted by $\STr$, 
requires that we symmetrize over all permutations of $\partial_a\Phi^i$ and
$[\Phi^i,\Phi^j]$ within the gauge trace upon expanding the square root.
For explicit computations, it is often convenient to write the two
determinants under the square root in terms of the single determinant,
\beq
S=-T_1\int d^2\sigma\,\STr\left[-\det\left(
\begin{array}{cc}
\eta_{ab}&\lambda\partial_a\Phi^j\\
-\lambda\partial_b\Phi^i&Q^{ij}
\end{array}
\right)\right]^{1\over2}.\labell{rewrite}
\eeq

To find static solutions describing the D3$\perp$D1-system, we allow three of
the transverse scalars, $\Phi^i$ ($i=1,2,3$), to depend on the spatial
coordinate $\sigma$. Evaluating the determinant in eq.~\reef{rewrite}
the energy becomes
\beqa
E&=&T_1\int d\sigma\,\STr\sqrt{1+\lambda^2(\partial_\sigma\Phi^i)^2
-{1\over2}\lambda^2[\Phi^i,\Phi^j]^2-{1\over4}\lambda^4(\epsilon^{ijk}
\partial_\sigma\Phi^i[\Phi^j,\Phi^k])^2}\nonumber\\
&=&T_1\int d\sigma\, \STr\sqrt{\lambda^2(\partial_\sigma\Phi^i
\mp{i\over2}\epsilon^{ijk}[\Phi^j,\Phi^k])^2+
(1\pm{i\over2}\lambda^2\epsilon^{ijk}\partial_\sigma\Phi^i
[\Phi^j,\Phi^k])^2}\nonumber\\
&\ge&T_1\int d\sigma\,\STr\left(1\pm{i\over2}\lambda^2\epsilon^{ijk}
\partial_\sigma\Phi^i
[\Phi^j,\Phi^k]\right)\nonumber\\
&=&NT_1\int d\sigma\pm{i\over3}\lambda^2 T_1\,\int d\sigma\,
\partial_\sigma\!\Tr\left(\epsilon^{ijk}\Phi^i\Phi^j\Phi^k\right).
\labell{d1d3enbound}
\eeqa
Here the lower
bound is again the sum of a trivial term (the energy of the $N$
D-strings) and a topological term. The minimum energy condition is
\beq
\partial_\sigma\Phi^i=\pm{i\over2}\epsilon^{ijk}[\Phi^j,\Phi^k],
\labell{nahm}
\eeq
which can be identified as the Nahm equations \cite{nahm,ded}. 
The desired solution is given by
\beq
\Phi^i(\sigma)=\pm{\alpha^i\over2\sigma},
\labell{Phisol}
\eeq
where the $\alpha^i$ are an $N\times N$ representation of the $SU(2)$
algebra, 
\beq
[\alpha^i,\alpha^j]=2i\epsilon^{ijk}\alpha^k.
\labell{su2alg}
\eeq
These $SU(2)$ matrices satisfy $\sum (\alpha^i)^2= C\,\identity_N$,
where $\identity_N$ the $N\times N$ identity matrix, and the
constant $C$ is the Casimir. We focus on the
irreducible $N\times N$ representation 
for which $C=N^2-1$. This noncommutative scalar field configuration
(\ref{Phisol}) describes a fuzzy two-sphere \cite{two} with a physical radius
\beq
R(\sigma)=\lambda\sqrt{{1\over N}\Tr[\Phi^i(\sigma)^2]}=
{\sqrt{C}\pi \ls^2\over\sigma}
={N\pi\ls^2\over\sigma}\sqrt{1-1/N^2}.
\labell{radstr}
\eeq
Hence the solution describes a ``fuzzy funnel'' in which the D-strings
expand to fill the $X^{1,2,3}$ hyperplane at $\sigma=0$.
This geometry can be compared to the D3-brane solution (\ref{phisol}) after
relabeling $\sigma\rightarrow\lambda\phi$ and $R\rightarrow r$. We see that
both descriptions yield the same geometry in the limit of large $N$,
up to $1/N^2$ corrections. There is similar agreement for other quantities,
such as the energy,
\beq
E=N T_1\int_0^\infty d\sigma+(1-1/N^2)^{-1/2}T_3\int 4\pi R^2 dR,
\labell{d1d3energy}
\eeq
and the charges, in the large
$N$ limit. Hence at least in this limit, the two dual descriptions
are in good agreement, as might be expected since
the fuzziness of the funnel solution is effectively smoothed out.
Note, however, that the Born-Infeld actions, eqs.~\reef{act3}
and \reef{action}, only describe low energy dynamics and do not
provide a complete theory in either case. Therefore, in general, one must
regard these two dual descriptions as complimentary. That is, the D-string
theory gives reliable results near the core of the spike, while
the D3-brane theory is valid for $r$ large --- see
ref.~\cite{bioncore} for further discussion of these issues.

\section{D-strings growing into D5-branes}

As we reviewed above and elaborated on
in ref.~\cite{bioncore}, the non-abelian theory describing $N$ coincident
D-strings has noncommutative scalar field solutions which have an
interesting interpretation in terms of fuzzy geometry. To describe
the D3$\perp$D1-system, it is natural to consider a
funnel with a fuzzy two-sphere as the cross-section.
However, this particular noncommutative geometry was introduced by hand,
and so a natural question to investigate is whether similar solutions of the
D-string equations of motion exist which involve fuzzy geometries other
than the two-sphere. In the following, we focus on a generalization
involving fuzzy four-spheres based on  
the matrices constructed by Castelino, Lee and Taylor \cite{wati4}.
As might be expected from their analysis, 
we will see that this construction leads to
a fuzzy funnel in which the D-strings expand into orthogonal D5-branes. 
Although not supersymmetric, this configuration shares many common
features with the D3-brane funnel. We will comment briefly on
other noncommutative geometries in the discussion section.

Our starting point is again the low energy action for $N$ D-strings
(\ref{action}), however, we now consider static configurations involving
five (rather than three) nontrivial scalars, 
$\Phi^i$ with $i=1,\ldots,5$. In this case, the action becomes
\beqa
S&=&-T_1\int d^2\sigma\,\STr\left\{
1+\lambda^2(\partial_\sigma\Phi^i)^2+2\lambda^2\Phi^{ij}\Phi^{ji}
+2\lambda^4(\Phi^{ij}\Phi^{ji})^2-4\lambda^4\Phi^{ij}\Phi^{jk}
\Phi^{kl}\Phi^{li}+\phantom{1\over4}\right.\nonumber\\
&&\!\!\left.
\phantom{1}+2\lambda^4(\partial_\sigma\Phi^i)^2\Phi^{jk}
\Phi^{kj}-4\lambda^4\partial_\sigma\Phi^i\Phi^{ij}\Phi^{jk}
\partial_\sigma\Phi^{k}+{\lambda^6\over4}(\epsilon^{ijklm}
\partial_\sigma\Phi^i\Phi^{jk}\Phi^{lm})^2\right\}^{1/2},
\labell{expact}
\eeqa
where we have introduced the convenient notation
\beq
\Phi^{ij}\equiv{1\over2}\left[\Phi^i,\Phi^j\right].
\labell{natation}
\eeq

To construct a new funnel solution, we consider the following ansatz:
\beq
\Phi^i(\sigma)=\pm\hR(\sigma)\,G^i,\ \ i=1,\ldots,5\ ,
\labell{ansatz}
\eeq
where $\hR(\sigma)$ is the (positive) radial profile
and $G^i$ are the matrices constructed in ref.~\cite{wati4}
--- see also ref.~\cite{grosse}. We give the definition and many useful
properties of the $G^i$ matrices in appendix A. 
Here, we simply note that
the $G^i$ are given by the totally symmetric $n$-fold tensor product
of $4\times4$ gamma matrices, and that the dimension of the matrices
is related to the integer $n$ by
\beq
N={(n+1)(n+2)(n+3)\over6}. \labell{Nnrel}
\eeq
The solution will describe a funnel whose cross-section is a
fuzzy four-sphere with a physical radius
\beq
R(\sigma)=\lambda\sqrt{\Tr[\Phi^i(\sigma)^2]/N}
=\lambda \hR(\sigma)\sqrt{\Tr G^i G^i/N}=
\sqrt{c}\lambda\hR(\sigma).
\labell{physR}
\eeq
The constant $c$ is the ``Casimir'' associated with the $G^i$
matrices, \ie $G^i G^i=c\,\identity_N$, given by
\beq
c=n(n+4).
\labell{kasi}
\eeq

Before considering the full solution, we remind the reader of
a puzzle presented in ref.~\cite{bioncore}. Consider the lowest
order equations of motion,
\beq
\left(-\prt^2_\tau+\prt_\sigma^2\right)\Phi^i=[\Phi^j,[\Phi^j,\Phi^i]],
\labell{motion}
\eeq
which are independent of the number of transverse scalars being
excited. Substituting in the static ansatz (\ref{ansatz}), and applying
various matrix identities presented in the appendix, yields
\beq
\hR''=16 \hR^3,
\labell{universe}
\eeq
which yields a simple funnel solution: $\hR(\sigma)=(2\sqrt{2}\sigma)^{-1}$. 
Now, this solution shows the same $R\sim\sigma^{-1}$ behavior
as we found for the D3-brane funnel (\ref{Phisol}). In fact, 
since the lowest order equations always take the same form,
this will be the universal small $R$ behavior of any D-string funnel.
However, if the funnel constructed here is to expand into a D5-brane,
we expect that at large $R$, the expansion would be governed by
a harmonic function in five spatial dimensions, \ie $\sigma\sim R^{-3}$
or $R\sim \sigma^{-1/3}$. In ref.~\cite{bioncore} we
anticipated the resolution of this puzzle, namely, that higher
order terms in the Born-Infeld action would effect a transition from
the universal small-$R$ behavior to the ``harmonic'' expansion at large $R$.
This is precisely the result that we find in the following.

Now consider the full equations of motion resulting from the action
\reef{expact}. It is a straightforward but tedious exercise to calculate the
latter, and the result is a rather formidable  and unilluminating
expression that we do not include here. 
However, plugging the
ansatz (\ref{ansatz}) into these equations, one finds that
all the matrix products simplify, leaving a single overall factor of $G^i$. To
verify this claim one uses the identities given in appendix A.
The final result is then simply a differential equation for $\hR(\sigma)$.

Knowing this, a simpler procedure is to substitute
the ansatz (\ref{ansatz}) directly into the action
(\ref{expact}). We again make heavy use of the matrix identities in
appendix A, and we choose to express the results in terms of the
physical radius $R$, rather than $\hR$, using eq.~(\ref{physR}).
The resulting action for the radial profile $R(\sigma)$ is
\beq
 S
=-NT_1\int d^2\sigma\sqrt{1+(R')^2}[1+4R^4/(c\lambda^2)].
\label{Raction}
\eeq
Actually this result is only captures the leading large-$N$ contribution at each
order in the expansion of the square root. There are corrections at order $1/c$
because we have not fully implemented the symmetrization of the trace.

From the action \reef{Raction}, we can derive an energy bound similar to
that in eq.~(\ref{d1d3enbound}), 
\beqa
E&=&NT_1\int d\sigma\left[\left(R'{\mp}\sqrt{8R^4/(c\lambda^2){+}16R^8/
(c\lambda^2)^2}\right)^2+
\left(1{\pm} R'\sqrt{8R^4/(c\lambda^2){+}16R^8/(c\lambda^2)^2}\right)^2
\right]^{1/2}
\nonumber\\
&\ge&NT_1\int d\sigma
\left(1\pm R'\sqrt{8R^4/(c\lambda^2)+16R^8/(c\lambda^2)^2}\right).
\labell{Renergy}
\eeqa
This is again a sum of the trivial term and a topological term.
The equality is obtained when
\beq
R'=\mp\sqrt{8R^4/(c\lambda^2)+16R^8/
(c\lambda^2)^2}.
\labell{Rpeq}
\eeq
Note that this equation is also compatible with the full equation of motion 
--- see end of the section for more discussion.

Let us now investigate the profile specified by eq.~(\ref{Rpeq}). 
This equation can be explicitly solved in terms of elliptic functions,
but it is more instructive to consider various limits.
For small $R$,  the $R^4$ term under the square
root dominates, and we find the funnel solution 
\beq
R(\sigma)\simeq{\sqrt{c}\lambda\over2\sqrt{2}\sigma}.
\labell{Rasymlarge}
\eeq
This is precisely the leading order solution found above with
the universal behavior: $R\sim\sigma^{-1}$.
However, for large $R$ the equation becomes
$R'=\mp4R^4/(c\lambda^2)$, with solution
\beq\labell{Rasymeq}
R(\sigma)\simeq\left({c\lambda^2\over12\sigma}\right)^{1/3},
\eeq
which is precisely the harmonic behavior that we anticipated for a 
D5-brane to appear at $\sigma=0$. The cross-over between the universal and
harmonic expansion occurs when the two terms under the square
root are comparable, {\it i.e.},
\beq
R_c\sim (c\lambda^2/2)^{1/4}=(2\pi^2c)^{1/4}\ls.
\eeq
Note that for large $c$ (and hence large $N$), this is a macroscopic
distance scale, \ie $R_c\gg\ls$.

To verify that the solution of eq.~(\ref{Rpeq}) indeed
corresponds to the funnel expanding into a number of D5-branes,
we can investigate the RR charge and the energy.
If the configuration corresponds to 
D5-branes located at $\sigma=0$ and spanning the $X^{1,2,3,4,5}$
directions, it should be a source for the six-form
RR potential $C^{(6)}_{012345}$. The
required source term comes from the following term in the non-abelian
Wess-Zumino action \cite{dielec,watiprep},
\beqa
-{\lambda^2\mu_1\over2}\int\STr\,P\left[(\hi_\Phi 
\hi_\Phi)^2 C^{(6)}\right]&=&
-{\lambda^3\mu_1\over2}\,\int d\sigma d\tau\,
C^{(6)}_{012345}\STr(\epsilon^{ijklm}\Phi^i \Phi^j\Phi^k\Phi^l
\partial_\sigma\Phi^m)\nonumber\\
&=&\pm{6N(n+2)\over c^{3/2}}\mu_5\int d\tau dR\ \Omega_4 R^4
\, C^{(6)}_{012345},
\labell{sourse}
\eeqa
where $\mu_5=\mu_1/(2\pi\ls)^4$ and 
$\Omega_4=8\pi^2/3$ is the area of a unit four-sphere. 
This is precisely the D5-source
term we expect, after averaging over the angles, and the
number of D5-branes is given by
\beq
{6N(n+2)\over c^{3/2}}={(n+1)(n+2)^2(n+3)\over n^{3/2}(n+4)^{3/2}}\simeq n,
\eeq
for  large $N$. Thus, in this limit, the funnel appears to
expand into $n$ D5-branes.
We can also now comment on the alternate signs which we introduced in the
ansatz \reef{ansatz}: With the plus sign, the funnel carries
a positive charge corresponding to $n$ D5-branes.
On the other hand, with the negative sign, there is a negative
charge corresponding to $n$ anti-D5-branes, by our present conventions.

Now consider the energy (\ref{Renergy}) of our configuration. 
Using eq.~(\ref{Rpeq}), this expression can be put into the following form,
\beqa
E&=&NT_1\int_0^\infty d\sigma\,[1+4R^4/(c\lambda^2)]^2\nonumber\\
&=&NT_1\int_0^\infty d\sigma+{6N\over c} T_5\int_0^\infty \Omega_4 R^4 dR
+NT_1\int_0^\infty dR - \Delta E,
\labell{energy}
\eeqa
where $T_5=T_1/(2\pi\ls)^4$.
The first term in eq.~(\ref{energy}) corresponds to
the energy of $N$ semi-infinite strings stretching from
$\sigma=0$ to infinity. The
second term gives the energy of $6N/c$ D5-branes spanning the
$X^{1,2,3,4,5}$ hyperplane. Note that in the large $N$ limit,
$6N/c\sim n$ and so the energy calculation reproduces the result that
$n$ D5-branes appear at $\sigma=0$, as we
inferred from the charge calculation above.
These first two terms in eq.~\reef{energy}
precisely parallel the two terms making up the total energy of the supersymmetric 
D3$\perp$D1 system. The current configuration however is not supersymmetric
and so there are other contributions to the energy.
The third term seems to
correspond to the energy of $N$ D-strings running out radially across
D5-brane world volume.
The last contribution is a finite binding energy:
\beq
\Delta E=2NT_1\left({c\lambda^2\over2}\right)^{1/4}
\int_0^\infty du u^4\left[1+{1\over2 u^4}-\sqrt{1+{1\over u^4}}\right]
\approx 1.0102 N c^{1/4}T_1 \ls\ .
\labell{binde}
\eeq
We will further discuss this result for the energy in the last section.

The above discussion concerns only an infinite funnel solution, but let us
return to the full equation of motion to look for other solutions.
The equation of motion derived from the action
(\ref{Raction}) for the radial profile is
\beq
{1\over R'}{d\over d\sigma}{1+4R^4/(c\lambda^2)\over
\sqrt{1+(R')^2}}=0.
\labell{fulleq}
\eeq
This is easily integrated once to yield
\beq\labell{fulleqsol}
(R')^2=B^2\left(1+{4R^4\over c\lambda^2}\right)^2-1,
\eeq
where $B^2$ is a dimensionless integration constant. 
Just as for the D3$\perp$D1 system
\cite{bioncore}, there are three classes of solutions depending 
on the value of $B^2$. If $B^2=1$, eq.~\reef{fulleqsol} coincides with
the condition \reef{Rpeq} minimizing the energy \reef{Renergy}.
Hence, this choice reproduces the infinite funnels discussed above.
For $B^2<1$, the solution reaches $R=0$ at finite
$\sigma$ yielding a pinched off funnel. This solution is naturally
continued beyond $R=0$ with another pinched funnel, and the resulting
configuration would have the interpretation of two sets of parallel
D5-branes joined by a finite D-string. Finally, for $B^2>1$, the
solution reaches $R'=0$ at finite $\sigma$ and terminates. Again 
by matching the appropriate solutions, we
can construct a double funnel which in this case describes a D-string
joining parallel D5-branes and anti-D5-branes. We do not elaborate
on the details of the double funnel constructions as they are made in
complete analogy with those in the D3$\perp$D1 system ---
see Section~3.2 of ref.~\cite{bioncore}.

\section{D5-branes growing D-strings}

We have seen in the previous section that the D-string theory contains
solutions where the D-strings expand into a set of orthogonal D5-branes.
Now we want to investigate to what extent these results can be reproduced
from the dual point of view of the D5-brane world volume theory.
The charge conservation arguments presented in the introduction showed
that the orthogonal D-strings act as a source of second Chern class in the
world volume theory of the D5-branes. To be precise, if $N$ D-strings end
on a set of D5-branes, we require
\beq
{1\over 8\pi^2}\int_{S^4}\Tr\, F\wedge F=N,
\labell{chargeeq}
\eeq
for any four-sphere surrounding the D-string endpoints. Hence we must excite
a non-abelian gauge field in the D5-brane theory, which in turn means that we
must consider a collection of, say, $n$ coincident D5-branes for which
the world volume gauge symmetry is enhanced to $U(n)$.
By analogy to the bion spike on the D3-brane,
we must also excite one of the transverse scalars, say $\phi=\phi^9$,
but it must reside in the overall 
$U(1)$ component of the $U(n)$ gauge symmetry to represent a collective
deformation of the geometry of all the D5-branes. The relevant 
world volume action for the $n$ D5-branes reads
\beq\labell{D5action}
S_5=-T_5\int d^6\!\sigma\ \STr\sqrt{-\det\left(G_{ab}+
\lambda^2\partial_a\phi\partial_b\phi+\lambda F_{ab}\right)}.
\eeq
We introduce spherical coordinates on the D5-brane world volume, with
radius $r$ and angles $\alpha^i\ (i=1,\ldots,4)$. The line element
may be written
\beq
ds^2=G_{ab}d\sigma^a d\sigma^b=-dt^2+dr^2+r^2 g_{ij}d\alpha^i d\alpha^j,
\eeq
where $g_{ij}$ is the metric on a four-sphere with unit radius,
\beq
g_{ij}={\rm diag}[1,\ \sin^2(\alpha^1),\ \sin^2(\alpha^1)\,\sin^2(\alpha^2),\ 
\sin^2(\alpha^1)\,\sin^2(\alpha^2)\,\sin^2(\alpha^3)]\ .
\labell{jdsuni}
\eeq
Now we look solutions with a ``nearly spherically symmetric'' ansatz: The
scalar is only a function of the radius, \ie $\phi=\phi(r)$.
For the gauge field, we require that $A_r=0$ while the angular components
are independent of $r$, \ie, $A_{\alpha^i}=A_{\alpha^i}(\alpha^j)$.
Examining the full equations of motion shows that this is a consistent ansatz.
For the non-vanishing components of the field strength, we
introduce the convenient notation: $\tF_{ij}\equiv\lambda
F_{\alpha^i\alpha^j}$. Further define
\beq
\tF^{ij}\equiv g^{ik}g^{jl}\tF_{kl}, \qquad\quad
\dtF_{ij}={1\over2}\epsilon_{ijkl}\tF^{kl},
\labell{convien}
\eeq
\ie the indices are raised with the inverse metric on the unit
four-sphere, and the dual is defined using the volume form of the
unit four-sphere, with $\epsilon_{1234}=\sqrt{g}$\ (where $g\equiv\det(g_{ij})$). With these
definitions, the charge conservation condition (\ref{chargeeq}) becomes
\beq
{1\over16\pi^2\lambda^2}\int \Tr\left(\tF_{ij}\dtF^{ij}\right)
\sqrt{g}\,d^4\alpha=N.
\labell{fwf2}
\eeq

Substituting in the ansatz given above, the action (\ref{D5action}) can be
evaluated as
\beq
S_5=-T_5\int d^6\!\sigma\,\sqrt{g}\,\sqrt{1+\lambda^2(\phi')^2}\,
\STr\sqrt{
r^8+{1\over2}r^4\tF_{ij}\tF^{ij}+{1\over16}(\tF_{ij}\dtF^{ij})^2}\ .
\labell{D5actexp}
\eeq
As in the previous section, the expression under the square root 
does not lend itself to being written as a sum of squares which would
facilitate a minimum-energy analysis.
We will return to this point later, but for now we study
instead the equations of motion directly.

The equation of motion for $\phi$ is simply
\beq
{d\over dr}{\partial S_5\over\partial\phi'}=0\ ,
\eeq
which yields 
\beq\label{phieq}
{\lambda^2\phi'\over\sqrt{1+\lambda^2(\phi')^2}}=
{f(\alpha^i)/\sqrt{g}\over
\STr \sqrt{r^8+{1\over2}
r^4\tF_{ij}\tF^{ij}+{1\over16}(\tF_{ij}\dtF^{ij})^2}}\ ,
\eeq
where $f(\alpha^i)$ is an arbitrary function independent of
$r$. Now the left hand side is a
function of $r$ only, so for consistency with our ansatz,
we require: $f(\alpha^i)=\lambda^3\sqrt{g}/\tB$ where $\tB$
is a dimensionless integration constant, and the gauge field must be
such that the denominator above is independent
of the angles. 

The gauge field equation of motion can be written
\beq
D_i\left[\sqrt{g}{r^4\tF^{ij}+{1\over4}\dtF^{ij}\tF_{kl}\dtF^{kl}
\over
\sqrt{r^8+{1\over2}r^4\tF_{ij}\tF^{ij}+{1\over16}(\tF_{ij}\dtF^{ij})^2}}
\right]=0,
\eeq
It is easy to see that if we choose a self-dual field strength
$\tF_{ij}=\dtF_{ij}$, the above equation reduces to the usual
Yang-Mills equations, which are then automatically satisfied due to the
Bianchi identity. The scalar equation of motion (\ref{phieq}) then imposes
the further restriction that 
$\tF_{ij}\dtF^{ij}$ should be independent of the angles, {\it i.e.},
the equations of motion restrict our ``nearly spherically symmetric''
ansatz to a fully spherically symmetric solution.
Of course, we could also choose an anti-self-dual field
strength with negative instanton number (representing an anti-D-string spike),
giving identical results.

For now let us assume that the desired instanton solutions exist
--- we will return to this point below. With
$\int\!\sqrt{g}\,d^4\alpha=8\pi^2/3$, eq.~(\ref{fwf2}) yields
that
\beq\labell{fdfeq}
\Tr\, \tF_{ij}\dtF^{ij}=6N\lambda^2.
\eeq
Similarly with a self-dual gauge field, the square root that appears in 
various expressions above becomes
\beq
\STr\sqrt{r^8+{1\over2}r^4\tF_{ij}\tF^{ij}+{1\over16}(\tF_{ij}\dtF^{ij})^2}
=\Tr\left(r^4+{1\over4}\tF_{ij}\dtF^{ij}\right)
=n r^4+{3\over2}N\lambda^2.\labell{strsqrt}
\eeq
The equation of motion (\ref{phieq}) for $\phi(r)$ can then be written
\beq
\lambda^2(\phi')^2={1\over\tB^2({n r^4\over\lambda^2}+{3\over2}N)^2-1}.
\labell{skale}
\eeq
To compare with the radial profile (\ref{fulleqsol}) 
found in the D-string description, we identify the physical
transverse distance as $\sigma=\lambda\phi$, and equate the radii
$r=R$. Then we see the form of the two equations agrees provided
that we set $\tB=2B/3N$. Complete agreement of the equations requires
the coefficients $4/c$ and $2n/(3N)$ to be equal, and in fact, this
equality is achieved in the large $N$ limit. Hence in this limit we have
complete agreement for the geometry determined by
the two dual approaches! This agreement would hold for the double funnel
solutions discussed below eq.~(\ref{fulleqsol}), as well as for the infinite
funnel which is the focus of our investigation.

For the spherically symmetric spike describing semi-infinite D-strings,
the energy  is easily evaluated to be
\beqa
E&=&T_5\int\!\!\sqrt{g}\,d^4\alpha\, dr\ \sqrt{1+\lambda^2(\phi')^2}
\,(nr^4+{3\over2}N\lambda^2)\nonumber\\
&=&N T_1\int d\sigma+n T_5\int \Omega_4 r^4dr+NT_1\int dr-\Delta E,
\labell{D5en}
\eeqa
with $\Delta E\approx1.0102 N(6N/n)^{1/4}T_1 \ls$. Again we have full
agreement with the D-string result (\ref{energy}) in the limit of large $N$. 
Note that in analogy with the D3$\perp$D1 analysis, the above
result includes a contribution of precisely $n$ D5-branes,
while the D-string calculations yield $1/n$ corrections to the
coefficient of this term.

The one point we have not yet addressed is finding an actual gauge field
solution corresponding to a homogeneous instanton on the 
four-sphere.\footnote{We would like to acknowledge Jan Segert for invaluable
assistance on this issue.}
The ADHM construction reduces the problem
of finding instanton configurations on $\bbR^4$ to an explicit algebraic
procedure --- see, \eg ref.~\cite{atiyah}. Due to conformal invariance
of the Yang-Mills system, instanton solutions on the four-sphere can then be
produced with the usual stereographic projection of $S^4$ onto $\bbR^4$.
For instanton number $N=1$ (or --1) and gauge group $SU(2)$, one finds that
the unit size instanton located at the origin of $\bbR^4$ projects to a 
homogeneous instanton configuration on the four-sphere.
We describe this configuration is some detail in appendix B for the
interested reader. Replacing the fundamental $SU(2)$ generators  by
an $n\times n$ representation allows us to embed this homogeneous solution
in an $SU(n)$ gauge theory. 
The instanton number of the resulting $SU(n)$ gauge
field is maximized by choosing the irreducible representation, giving 
\beq
N={1\over 8\pi^2}\int_{S^4}\Tr\, F\wedge F={n(n^2-1)\over6}.
\labell{suninst}
\eeq
Now a key result is that
using a theorem by H.C.~Wang \cite{wang}, one may prove that this is the {\it
homogeneous} instanton configuration on the four-sphere with the largest
possible value of the second Chern class --- see discussion in Appendix B.
So in particular for
$SU(2)$ instantons on $S^4$, the only $SO(5)$ invariant configurations
will have $N=1,0$ or --1. 
For the higher rank case of $SU(n)$, we can of course obtain
an instanton number lower than in eq.~\reef{suninst} by choosing a reducible 
$n\times n$ representation of the $SU(2)$ algebra. This would correspond to 
several decoupled bunches of D5-branes each of which would support separate
bundles of D-strings. Also, the scalar
field $\phi(r)$ would necessarily be a more general diagonal matrix made up of 
a direct sum of independent U(1) subgroups rather than
proportional to the full $n\times n$ identity matrix. In this situation  
the different parts of the problem completely decouple.  

In fact we see from the energy (\ref{D5en}) that the irreducible
representation is preferred: As we will describe in the discussion
section, for a given number $N$ of D-strings and
an unlimited supply of D5-branes, we maximize $\Delta E$ (and thus
minimize the energy) by coupling the D-strings as a single bundle to a
number of D5-branes given by eq. (\ref{suninst}). 

Note that for large $n$, the upper bound 
(\ref{suninst}) on $N$ agrees exactly
with the mysterious restriction (\ref{Nnrel}) that appears in the
construction of the fuzzy funnel! While this agreement is a remarkable
success of the duality between the D-string and D5-brane descriptions,
it appears that this restriction arises from our use of a spherically
symmetric ansatz. Presumably the restriction can be evaded if we were
to consider more general field configurations, although as a practical
matter this would make the analysis much more difficult. 

With a fully spherically symmetric ansatz as required by the equations
of motion, we can also perform a minimum-energy analysis in analogy
with the D3$\perp$D1-brane case.
Beginning with D5-brane action (\ref{D5actexp}), the energy
may be suggestively written as
\beqa
E&=&T_5\int\!\!\sqrt{g}\,d^4\alpha\,r^4dr\,\STr\left[
(\sqrt{1+\textstyle{1\over2}r^{-4}\tF\dtF}\lambda\phi'\mp
\textstyle{1\over4}r^{-4}\tF\dtF)^2+\right.\nonumber\\
&&\!\!\left.\phantom{1}+
\textstyle{1\over4}r^{-4}(\tF-\dtF)^2(1+\lambda^2(\phi')^2)
+(\sqrt{1+\textstyle{1\over2}r^{-4}\tF\dtF}\pm
\textstyle{1\over4}r^{-4}\tF\dtF\lambda\phi')^2\right]^{1/2}\nonumber\\
&\ge&T_5\int\!\!\sqrt{g}\,d^4\alpha\,r^4dr\,\Tr\left[
\sqrt{1+\textstyle{1\over2}r^{-4}\tF\dtF}\pm
\textstyle{1\over4}r^{-4}\tF\dtF\lambda\phi'\right],
\labell{D5enbound}
\eeqa
where $\tF\dtF\equiv\tF_{ij}\dtF^{ij}$. 
The lower bound on the energy is achieved\footnote{There is a similar bound
for anti-self-dual field strengths, obtained by flipping the sign of $\dtF$
everywhere.} with $\tF_{ij}=\dtF_{ij}$ and
\beq
\lambda\phi'=\pm{1\over4}r^{-4}\tF\dtF/\sqrt{1+\textstyle{1\over2}r^{-4}
\tF\dtF}.
\labell{minhum}
\eeq
Now substituting in eq.~\reef{fdfeq}, eq.~\reef{minhum} 
coincides precisely with the equation governing the profile
of the infinite spike, \ie eq.~\reef{skale} with $\tB=2/3N$.
Claiming that the lower bound is a true minimum energy
is not quite rigorous, as the lower bound is not just a sum of ``trivial'' and
``topological'' terms, as in the D3$\perp$D1-brane case. Combining
spherical symmetry and the charge quantization condition (\ref{fwf2}),
one has that $\tF\dtF$ is independent of $r$, and so the second term is indeed
topological. Note that the ``trivial'' term contains the contributions
of the D5-branes, the radial D-strings and the binding energy.

Note that  for eq.~\reef{minhum} to have a sensible interpretation,
all of the gauge covariant expressions are understood to be
proportional to the identity matrix. This is achieved by selecting the 
irreducible $SU(2)$ representation in the construction of the
instanton as described above. For the reducible cases, the 
equation (\ref{D5enbound})
decouples into separate parts after extending $\phi(r)$ to a
more general diagonal matrix, again as discussed above.

\section{Generalization to dyonic strings}

It is straightforward to extend the discussion to dyonic strings
\cite{bound,jhs,one}, rather than D-strings, ending on D5-branes.
That is, replace the D-strings above with $(N,N_f)$-strings, \ie
bound states of $N$ D-strings and $N_f$ fundamental strings.
The revised analysis is extended in complete analogy with that
for the D3$\perp$D1 system -- see, e.g., refs.~\cite{bioncore,gauntlett} --
and so we present only the salient calculations.

Let us start with the D-string world volume theory, where fundamental strings
are introduced by adding  a $U(1)$ electric field.  Denoting the electric field
as $F_{\tau\sigma}={\cal E}\,\identity_N$, the D-string action (\ref{Raction})
becomes
\beq
S=-NT_1\int d^2\sigma\sqrt{
1-\lambda^2{\cal E}^2+(R')^2}[1+4R^4/(c\lambda^2)].
\labell{neu}
\eeq
Now the gauge field equations of motion set $\cal E$ to be a constant.
The latter is then fixed by the quantization condition on the 
displacement, $D=N_f/N$, where
\beq
D\equiv{1\over N}{\delta S\over\delta {\cal E}}=
{\lambda^2 T_1{\cal E}\over\sqrt{1-\lambda^2{\cal E}^2}},
\eeq
after using the equations of motion.
We note that rescaling $\sigma$ to $\hat{\sigma}=\sqrt{1-\lambda^2{\cal E}^2}
\,\sigma$ in eq.~\reef{neu} precisely reproduces the action \reef{Raction}
with no electric field. Hence the modified profile, 
in the presence of the $N_f$  fundamental strings, is 
just a stretched out version of the previous result. 
To be precise, we have
\beq
\sigma_{(N,N_f)}(R)=\sqrt{1+(g_sN_f/N)^2}\,\sigma_{(N,0)}(R),
\labell{sigmap2}
\eeq
using $T_1=(\lambda g_s)^{-1}$ where $g_s$ is the string
coupling. Hence for large $N_f$, the profile keeps behaving like the harmonic
solution further out from the D5-brane, as would be
expected for a collection of fundamental strings --- 
see the discussion section.
To calculate the energy, we
evaluate the Hamiltonian, $E=\int d\sigma(D{\cal E}-{\cal L})$. The only
difference in the final result compared to eq.~(\ref{energy}),
is that the contribution
involving the perpendicular strings becomes
\beq
E\simeq \sqrt{N^2+g_s^2 N_f^2}\,T_1\int_0^\infty d\sigma,
\labell{edyon}
\eeq
as expected for the dyonic strings. The energy of the radial
D-strings, however, is unchanged, in accord with expectations.

From the point of view of the D5-brane world volume theory, the
fundamental strings act as electric point charges \cite{calmaldetc}.
Hence in this case,  we add a
static radial electric field ${\cal E}(r)=F_{0r}(r)=-A_0'(r)$ in 
the $U(1)$ sector of the D5-brane theory. The only change in 
the D5-brane action (\ref{D5actexp})
is that $(\phi')^2\rightarrow(\phi')^2-(A_0')^2$, which means that
$\phi(r)$ and $A_0(r)$ will have the same equations of motion. Therefore
we will find solutions with $A_0(r)=\alpha\phi(r)$, where $\alpha$ a
constant. In
terms of the rescaled scalar, $\hat{\phi}(r)=\sqrt{1-\alpha^2}\phi(r)$
the equations and the solution are exactly as before. 
Again the only modification to the energy is that the contribution
representing the perpendicular strings is
multiplied by a factor of $1/\sqrt{1-\alpha^2}$. We can use
this result to read off the value of $\alpha$, namely,
\beq
\alpha={g_sN_f\over\sqrt{N^2+g_s^2N_f^2}},
\eeq
completing the specification the solution. This constant can also
be fixed to the same value by charge quantization arguments.
In any event, this value of $\alpha$ produces precisely the same
stretching of the radial profile as was found in the D-string 
analysis \reef{sigmap2}. Notice that the
electric field associated with the fundamental strings is not just a simple
Coloumb field, rather it has the structure of the profile function
$\phi'(r)$. For large $g_sN_f$, however, the region where the field is
approximately a Coloumb field becomes large. Further in this limit,
the field agrees precisely with that expected for an fundamental-string spike
\reef{fspike}.

\section{Linearized fluctuations}

For the D3$\perp$D1 system, the dynamics of the both the bion spike
\cite{calmaldetc,waves}
and the fuzzy funnel \cite{bioncore} were studied by considering
linearized fluctuations around the static solutions presented
in Section~2. Given the fluctuation equations, two interesting comparisons 
were made. First, one can compare the propagation of perturbations
on the D3-brane spike with that on the D-string funnel. Here, at least
for large $N$ and low angular momenta (on the two-sphere
cross-sections), 
there was good agreement between the two descriptions.
As explained in ref.~\cite{bioncore}, producing agreement for high angular
momentum would require an analysis which goes beyond the low energy
Born-Infeld action. A second interesting comparison was made to perturbations
propagating on a test string in the supergravity
background of an orthogonal D3-brane \cite{waves}. One finds that the
fluctuation equations for the bion correspond precisely to the wave
equations for the test string, including the modifications due to
the curved space time background. From a modern point of view, one 
can attribute~\cite{ads2} the 
success of the latter agreement as due to the AdS/CFT correspondence
\cite{ads1}.

In the following, we begin an examination of propagation of fluctuations
on the fuzzy funnel presented in Section~3. Our analysis is incomplete
in that we do not present a complete understanding of how our results
should be mapped to either those for a D5-brane bion spike or a test
D-string in a supergravity background.

When considering fluctuations of the D5$\perp$D1 system,
the setup similar to the D3$\perp$D1-case \cite{waves,bioncore}. 
There are two basic types of
funnel fluctuations,
the overall transverse modes in the directions perpendicular to both
the D5-branes and the D-strings (\ie $X^{6,7,8}$), and the relative
transverse modes which are transverse to the D-strings, but parallel to the
D5-brane world volume (\ie along 
$X^{1,2,3,4,5}$). For simplicity in the following, we will only
present examples of the overall transverse fluctuations.

\noindent{\bf Modes on the fuzzy four-sphere:}

For the moment, consider the general case of D-strings expanding into a
D($2k$+1)-brane. The associated funnel would naturally have the topology
of $R\times S^{2k}$. The perturbations of this geometry are naturally
decomposed in terms of spherical harmonics on the $2k$-sphere
--- in particular this seems natural if one is to compare to the
fluctuations of the dual bion spike in the D($2k$+1)-brane description.
The latter can be thought of as symmetric traceless tensor representations
of the $SO(2k+1)$ symmetry acting on the $S^{2k}$. To be precise, given
Cartesian coordinates $X^i$ in a ($2k$+1)-dimensional embedding space, the
$2k$-sphere can be described as
\beq
\sum_i (X^i)^2=c\ .
\labell{sphear}
\eeq
With this restriction, the spherical harmonics may then be written as
\beq
a_{i_1i_2\cdots i_\ell}\,X^{i_1}X^{i_2}\cdots X^{i_\ell}
\labell{harms}
\eeq
where the $a$ are naturally symmetric, because of the commuting nature
of the coordinates, and traceless, because of the restriction \reef{sphear}.
In the construction of a fuzzy sphere, one replaces the embedding
coordinates $X^i$ by finite dimensional matrices $G^i$ satisfying
\beq
\sum_i (G^i)^2=c\, \identity\ .
\labell{sphear2}
\eeq
The spherical harmonics are then replaced by
\beq
a_{i_1i_2\cdots i_\ell}\,G^{i_1}G^{i_2}\cdots G^{i_\ell}
\labell{harms2}
\eeq
with the $a$ defined as above. These matrix harmonics
can be regarded as yielding a modification of the algebra of the
functions on the sphere, similar to usual discussions of noncommutative
geometry \cite{old2}. However, since the products in eq.~\reef{harms2} 
will only yield
a finite number of linearly independent matrices, there will be an
upper bound on $\ell$. Thus the matrix construction
truncates the full algebra of functions on the sphere to those
with $\ell\le\ell_{max}$. Thus the star product on the fuzzy sphere
differs from that obtained by the deformation quantization of the Poisson
structure on the embedding space, \ie the latter is defined for the space of all
square integrable functions on the sphere \cite{kiril,hawkins}. 

Consider the specific case of $k=1$, \ie D-strings expanding into a
D3-brane with a funnel of topology $R\times S^{2}$. The fuzzy two-sphere
\cite{two} is constructed with the $G^i$ chosen to correspond to the $SU(2)$
generators $\alpha^i$, appearing in Section~2. As discussed
there, for the irreducible $N\times N$ representation, these matrices
satisfy
eq.~\reef{sphear2} with $c=N^2-1$. Because the $\alpha^i$ form a Lie algebra
\reef{su2alg}, the symmetrized products \reef{harms2} with $\ell\le N-1$ yield
all of the possible independent matrix products. Hence there is a precise
correspondence between these matrix expressions \reef{harms2} and the
spherical harmonics on $S^2$ up to $\ell_{max}=N-1$.

The construction of the fuzzy four-sphere \cite{wati4,grosse}, which
corresponds to $k=2$ above, appears very similar, but in fact it
yields a very different object. Here the appropriate $G^i$ are the $N \times N$
matrices described in appendix A. With these, one can construct matrix
harmonics \reef{harms2} with $\ell\le\ell_{max}=n$ \cite{grosse}. 
However, a key difference from the fuzzy two-sphere is that here the
$G^i$ do not form a Lie algebra, and as a result the algebra of these
matrix harmonics does not close \cite{wati4}! This can be understood by
noting that while the $G^i$ alone do not form a Lie algebra, the combination
of the $G^i$ and $G^{ij}$ do, giving a representation of the algebra $SO(1,5)$.
This can be seen from the definition of the $G^{ij}\equiv[G^i,G^j]/2$
and the commutators in eq.~\reef{comme}. Hence a closed algebra of
matrix functions would be given by
\beq
\tilde{a}_{a_1a_2\cdots a_\ell}\,\tG^{a_1}\tG^{a_2}\cdots \tG^{a_\ell}
\labell{harms3}
\eeq
where the $\tG^a$ are generators of $SO(1,5)$ with $a=1,\ldots,15$,
and the $\tilde{a}$ are naturally symmetric in the $SO(1,5)$ indices.
There are also a number of trace constraints arising
from the matrix identities given in 
eqs.~(\ref{gicas},\ref{idum},\ref{GGGGeq}). Having identified $\tG^a=G^a$ for $a=1\ldots5$, the
desired matrix harmonics would correspond to the subset of $\tilde{a}$
with nonvanishing entries only for indices $a_i\le5$. Thus while the fuzzy
four-sphere construction introduces an algebra that contains a truncated
set of the spherical harmonics on $S^4$, the algebra also contains a
large number of elements transforming under other representations of
the $SO(5)$ symmetry group that acts on the four-sphere. We refer the
reader to ref.~\cite{grosse} for
a precise description of the complete algebra in terms of representations
of $SO(5)$ (or rather $Spin(5)=Sp(4)$).

When it comes to the physical fluctuations of the fuzzy funnel describing
D-strings expanding into D5-branes, we stress that both sets of these modes
play a role. For the purposes of illustration, it is sufficient to only 
consider the lowest order equations of motion \reef{motion} and 
examine linearized fluctuations $\delta\Phi^m$. 
The linearized equation  is
\beq
(-\partial_{\tau}^2+\partial_{\sigma}^2)\delta\Phi^m \,=\,
[\Phi^j,[\Phi^j,\delta\Phi^m]]\,+\,[\Phi^j,[\delta\Phi^j,\Phi^m]]\,+\,
[\delta\Phi^j,[\Phi^j,\Phi^m]]\  .
\labell{fluctuate}
\eeq
As mentioned above, we will focus on overall transverse fluctuations
involving, {\it e.g.,} $\delta\Phi^6$. First consider the matrix harmonic 
modes \reef{harms2}, \ie
a fluctuation proportional to a symmetric traceless product of $G^i$'s,
\beq
\delta\Phi^6=a_{i_1i_2\cdots i_l}G^{i_1}G^{i_2}\cdots G^{i_l}.
\labell{phisix}
\eeq
The lowest order equation \reef{fluctuate} then becomes
\beq
\left[\partial_{\tau}^2-\partial_{\sigma}^2
+{l(l+3)\over2\sigma^2}\right]a_{i_1\cdots i_l}(\sigma,\tau)
=0.
\eeq
Notice that the double commutator terms on the right hand side of
eq.~\reef{fluctuate} have produced the correct angular momentum barrier
for a spherical harmonic mode propagating on the $S^4$. 

As discussed earlier, however, we need also consider more general
fluctuations which include the $G^{ij}$. The simplest example is given by,
$\delta\Phi^6=\tilde{a}_{ij}G^{ij}$ where $\tilde{a}_{ij}=-\tilde{a}_{ji}$.
Substituting into the equation of motion \reef{fluctuate} yields
\beq
\left[\partial_{\tau}^2-\partial_{\sigma}^2
+{1\over\sigma^2}\right]\tilde{a}_{ij}(\sigma,\tau)
=0.
\labell{overallt}
\eeq
Hence the equation of motion has essentially the same form as for
the matrix harmonics. This illustrates then that within the context of
the fuzzy funnel, the fluctuations involving $G^{ij}$ are equally 
valid physical modes.

If we were to compare the above fluctuations to those in the D5-brane
description of the intersection, it would seem that the 
$\delta\Phi^6$ fluctuations proportional to symmetric traceless products
of $G^i$ must correspond to $U(1)$ scalar field fluctuations of $\phi^6$
proportional the spherical harmonics on the (commutative) four-sphere.
Identifying the D5-brane modes corresponding to $\delta\Phi^6$ fluctuations
involving $G^{ij}$ is more puzzling. However, it seems that they
must be related to non-abelian excitations of the D5-brane theory.
It would be interesting to investigate this correspondence in detail, as it
may yield new insights on thinking about the fuzzy four-sphere geometry. 
Beyond the linearized equations of motion, clues as to the interpretation 
of these modes can be found by investigating the nontrivial couplings
which they induce. For example, the Wess-Zumino part of the non-abelian
D-string action contains a term \cite{dielec}:
\beq
-i{\lambda\mu_1\over3}\int d\sigma d\tau 
\Tr\left(\Phi^i\Phi^j\Phi^k\right)F^{(5)}_{\tau\sigma ijk},
\labell{D3charge}
\eeq
corresponding to a D3-brane dipole coupling. Hence when the
antisymmetric mode considered above propagates on the fuzzy
funnel, one finds 
\beq
-i\lambda\mu_1\int d\sigma d\tau 
\,\Tr(\Phi^i\Phi^j\delta\Phi^6)\,F^{(5)}_{\tau\sigma ij6}
=i{4\mu_1\over5c\lambda}\int d\sigma d\tau\,R^2(\sigma)\,\tilde{a}_{ij}
\,F^{(5)}_{\tau\sigma ij6}.
\labell{coupling}
\eeq
using the background $\Phi^i=R(\sigma)G^i/\sqrt{c}\lambda$.
We therefore see that perturbing by these modes induces a D3-brane dipole along
the fuzzy funnel. This coupling would vanish for any of the usual matrix harmonics 
given in eq.~\reef{phisix}.

\noindent{\bf Supergravity comparison:}

For the D3$\perp$D1 system, it was found that the linearized equations
for the full Born-Infeld action encoded some of the modifications of
the supergravity background generated by the D3-branes \cite{waves,bioncore}.
Here we consider the analogous analysis for the fuzzy funnel
describing the D5$\perp$D1 system.

The simplest modes are those which do not excite
internal modes on the $S^4$, \ie fluctuations
proportional to the identity matrix, say
\beq
\delta \Phi^6(\sigma,\tau)=f(\sigma,\tau)\identity_N\ .\labell{idfluct}
\eeq
Plugging this fluctuation into the full D-string action, 
together with the funnel ansatz $\Phi^i={R}G^i/\sqrt{c}\lambda$, we find
\beqa\labell{faction}
S&=&-N T_1\int d^2\sigma\sqrt{(1+R'^2)(1-\lambda^2\dot{f}^2)+
\lambda^2 f'^2}\left[1+4 R^4/(c\lambda^2)\right]\nonumber\\
&=&-N T_1\int d^2\sigma\, H\sqrt{1-\lambda^2\dot{f}^2+{\lambda^2\over H}
f'^2}\\
&=&-NT_1\int d^2\sigma\,\left(H-{1\over2}\lambda^2 H \dot{f}^2+
{1\over2}\lambda^2 f'^2+\ldots\right),\nonumber
\eeqa
with
\beq
H(\sigma)=\left(1+{4R(\sigma)^4\over c\lambda^2}\right)^2.
\eeq
In the second line above, we used the equation of motion for
$R(\sigma)$, while in the third line retained only quadratic terms in $f$,
as is appropriate to determine the linearized equations of motion. The
latter become
\beq
\left(H\partial_\tau^2-\partial_\sigma^2\right)f(\sigma,\tau)=0.
\labell{feom}
\eeq
Evaluating this result for large $\sigma$, using
the asymptotics (\ref{Rasymlarge}), yields
\beq
\left(\left[1+{n^2\lambda^2\over8\sigma^4}\right]\partial_\tau^2-
\partial_\sigma^2\right)f(\sigma,\tau)\simeq 0.
\labell{feom2}
\eeq
Here we have also used $c\simeq n^2$ for large $n$.

Now we wish to compare this result with the fluctuation
equation for a test D-string in some supergravity background, \eg
a D5-brane background. Assuming a diagonal background
metric $G_{\mu\nu}$, the test D-string action becomes
\beqa
\tilde{S}&=&-N T_1\int d^2\sigma\,
e^{-\phi}\sqrt{G_{\tau\tau}G_{\sigma\sigma}-
\lambda^2 G_{\sigma\sigma}G_{66}\dot{f}^2+
\lambda^2 G_{\tau\tau}G_{66}f'^2}\nonumber\\
&=&-N T_1\int d^2\sigma\, e^{-\phi}\sqrt{G_{\tau\tau} G_{\sigma\sigma}}
\left(1-{1\over2}\lambda^2{G_{66}\over
G_{\tau\tau}}\dot{f}^2+
{1\over2}\lambda^2{G_{66}\over G_{\sigma\sigma}}f'^2
\right)\\
&=&-N T_1\int d^2\sigma
\left(\sqrt{h}-{1\over2}\lambda^2 h^{3/2}\dot{f}^2+{1\over2}\lambda^2
\sqrt{h}f'^2\right).\nonumber
\eeqa
The second line above holds for arbitrary (diagonal) metrics, while the 
third line is evaluated for the metric of $n$ D5-branes,
\beqa
ds^2&=&h(\sigma)^{-1/2}\,\eta_{\mu\nu}dx^\mu dx^\nu+h(\sigma)^{1/2}(d\sigma^2
+\sigma^2 d\Omega_3^2),\nonumber\\
e^{-2\phi}&=&h(\sigma),\labell{D5backg}\\
h(\sigma)&=&1+L^2/\sigma^2,\ \ L^2=n\alpha'g_s,\nonumber
\eeqa
Note that the background RR field and the Wess-Zumino interactions do not
contribute to the quadratic action in this case. Hence the
the linearized equation of motion for $f$ is
\beq
(h\partial_\tau^2-\partial_\sigma^2+{L^2\over h\sigma^3}
\partial_\sigma)
f(\sigma,\tau)=0,
\labell{moti1}
\eeq
which does not appear to match the previous equation of motion
(\ref{feom}). One can improve on matching the form of the equations
by changing the spatial variable, $\tilde{\sigma}^2=\sigma^2+L^2$,
which yields
\beq
(h^2\partial_\tau^2-\partial_{\tilde{\sigma}}^2)
f(\tilde{\sigma},\tau)=0.
\labell{moti2}
\eeq
However there is still a mismatch because the `harmonic' functions
multiplying $\partial_\tau^2$ fails to agree.

The $1/\sigma^4$ terms in eq.~\reef{feom2} would be more appropriate
for a 3-brane-like background. This is perhaps not surprising as the
large $\sigma$ behavior of the D-string funnel is universal, and will
thus agree with what we found for D3$\perp$D1. The disagreement is
still puzzling, however, as we would expect both the D-string
and the test-string descriptions to be valid for large $\sigma$. 
If we insist that the quadratic action (\ref{faction}) should be
exactly reproduced
by a supergravity background, the latter must satisfy
\beq
G_{\tau\tau}=G_{66}=H^{-1}G_{\sigma\sigma}=H^{1/2}e^{\phi}.
\labell{bakk}
\eeq
These precise relations need only apply along the radial
line where the D-string is placed.
It would be very interesting to find a physical interpretation of such 
a background.

Note that for the F-string spikes discussed in ref.~\cite{waves},  
agreement with supergravity was only found for D3- and D4-branes. In 
the case of D5-branes one can argue that the bion description of the F-string 
spike has broken down at large $\sigma$. In contrast, we expect that
large $\sigma$ is precisely the regime where
the fuzzy funnel description of the D5$\perp$D1 system is reliable.

\section{Discussion}

We have described two dual descriptions of the D5$\perp$D1 system.
In particular, we have found solutions of both the low energy D1-brane
and D5-brane theories describing a set of $N$ D1-branes ending on 
a collection of $n$ orthogonal D5-branes. Good agreement was found
in the formulae describing the energy, RR couplings and the funnel
geometry in the two theories. While one may regard the two different types
of D-branes as giving dual descriptions of the same intersection, we stress
that our analysis is limited to the {\it low energy} world volume theory
in each case. Hence in neither case do we have a complete description
as in both theories the configurations of interest contain singularities.
Our analysis therefore yields two complementary descriptions of the
D5$\perp$D1 intersection. In the region far away from the D5-branes, the
fuzzy funnel construction of the D1-brane theory is reliable,
while near the D5-brane, the D5-brane bion picture is trustworthy. 

\noindent{\bf Regimes of validity:}

One is left then with the question of why do these two descriptions agree
so well for large $n$ or $N$. A similar agreement exists for the
D3$\perp$D1 system, and in that case
a careful analysis of the two complementary
theories showed that for large $N$ there is a large region of overlap in which
both theories give a reliable description of the intersection \cite{bioncore}.
A similar result is found upon examining the regimes of validity of the
D-string and D5-brane theories in the present case.
First, we must determine when we can confidently ignore the
higher derivative corrections to the Born-Infeld action arising from the usual
$\alpha'$ expansion of low energy string theory. Schematically we require that
$\ls |\partial^2\Phi|\ll|\partial\Phi|$. For the spike solution on the D5-brane,
this translates into $r^2\gg\ls^2$, while for the fuzzy funnel in the D-string
theory, we find that this restriction is equivalent to $r^2\ll c\,\ls^2$. So for
large $n$ with $c\simeq n^2$, there is a significant overlap region where
the higher derivatives are negligible in both theories and we might expect
good agreement.

As well as higher derivative corrections, the non-abelian action with
the symmetric trace prescription \cite{yet} requires additional
higher order commutator corrections \cite{notyet,bain,unpub}
--- see also the discussion in ref.~\cite{dielec}. For the fuzzy funnel, we
might require that $\ls^2|\hR'|\ll 1$ or $\ls^2\hR^2\ll 1$ since these
dimensionless  quantities characterize the relevant scalar field commutators.
In terms of the physical radius, both of these restrictions can be translated
into $r^2\ll c\,\ls^2$, which coincides with that found above. A more
conservative bound for avoiding higher commutator corrections is demanding that
the Taylor expansion of the square root in the action \reef{expact} should
converge rapidly. The latter produces the more restrictive condition that
$r^2\ll c^{1/2}\ls^2$. However, for large $n$, this still seems to leave room
for a significant region of overlap with the D5-brane description. 

The D5-brane theory is also non-abelian, and so one might consider the effect
of higher commutator corrections to the spike solution as well. However, we
argue that in fact these additional commutators do not affect
the configuration studied here. First, as can be seen in eq.~\reef{D5actexp},
the determinant in the five-brane action \reef{D5action} factors into two
separate terms: a radial term involving the abelian scalar, and
an angular term involving the non-abelian gauge field. Hence any higher
commutators will only affect the latter contribution. However, here we note
that the gauge field configuration of interest is self-dual (or
anti-self-dual), and investigations of the non-abelian Born-Infeld theory
indicate that the higher commutator corrections do not
affect such configurations. In particular, one finds that the spectrum of the
non-abelian Born-Infeld theory precisely matches that of the full string theory
when considering excitations around a self-dual field configuration
\cite{notyet,bain,unpub}. Hence it would appear that higher order
commutators will not modify the analysis in the D5-brane theory.

Hence from our analysis, we are led to conclude that in the regime
$\ls^2\ll r^2\ll n\,\ls^2$ both the D-string and D5-brane descriptions should be
reliable. Thus for large $n$, we have a large region of overlap and we should
expect good agreement between the D5-brane spike and the D-string's fuzzy
funnel. In Section~$3$, the calculations are also expected to have $1/c$
corrections since we did not implement the symmetrization of the gauge trace.
However, these corrections to the D-string analysis will be negligible as long as 
$n$ is large. This point was not a problem in the D5-brane analysis
since for (anti-) self dual gauge fields the square root
in the action simplifies, as shown in eq.~(\ref{strsqrt}). 
In passing, we also point out that our calculations neglect gravitational
effects, which is justified when $g_sN\ll 1$. However, we can simply
consider very weak string coupling since $g_s$ does not appear
in our calculations.

\noindent{\bf Geometric profile:}

The full nonlinear form of the Born-Infeld action was important 
to produce a funnel which made a transition between the
universal behavior ($R\sim\sigma^{-1}$) at small $R$ and the
harmonic behavior ($R\sim\sigma^{-1/3}$) at large $R$, as discussed
in Section~3. Of course, the same is true for the profile
of the bion spikes constructed in Section~4. 

\begin{figure}[htb]
\centerline{\includegraphics[width=8cm]{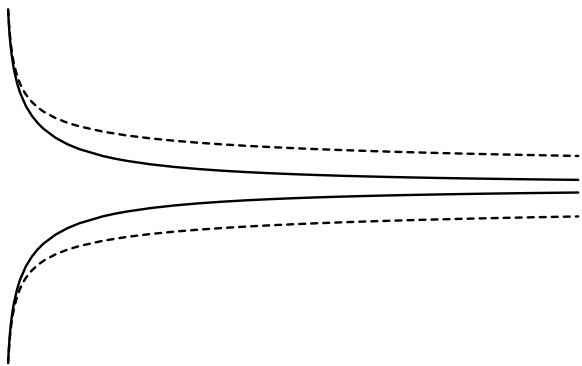}}
\caption{
The shape of D-string (solid curve) and fundamental string (dashed curve)
spikes extending from a set of orthogonal D5-branes. We have set $N=N_fg_s$
so that the profiles overlap for small $\sigma$. 
}\label{curves}
\end{figure}

It is interesting to compare the profile of the D5$\perp$D1 intersection
to that for the D5$\perp$F1 intersection. Fundamental strings ending
on an orthogonal D5-brane act like electric point sources in the world volume
theory of the D5-brane, and the intersection was studied from this point
of view in ref.~\cite{calmaldetc}. For $N_f$ fundamental
strings ending on $n$ D5-branes, the transverse scalar on the D5-brane
is given by
\beq
\sigma(r)={N_fg_s\lambda^2\over2n\,r^3},
\labell{fspike}
\eeq
and the corresponding radial electric field on the D5-brane world volume is
simply $E_r=\partial_r\sigma(r)$. This agrees with
the profile of the D-string spike for small $\sigma$ (\ie in the harmonic
function regime) with the substitution $N\rightarrow N_fg_s$. However,
for large $\sigma$, the D-string spike is much narrower because of the
onset of the universal behavior. The two profiles are plotted together
in Fig.~\ref{curves}.

S-duality will transform the D5$\perp$F1 intersection into an
NS5$\perp$D1 intersection. Since $g_s\ell_s^4\sim \ell_p^4$ is
invariant under S-duality, eq.~(\ref{fspike}) also describes
the profile for the latter system of intersecting branes.
Of course, this observation
begs the question of how does one realize this intersection with a fuzzy
funnel description? The latter is related to the unresolved problem of
constructing transverse five-branes in Matrix theory \cite{matrix,matrix2}.
For the profile (\ref{fspike}) above to be reliable in the D-string
theory, it would have to give a macroscopic radius in a regime where
the variation of the radius is still small (\ie $r\gg\ls,\,\partial_\sigma
r\ll 1$). Combining these constraints requires that $N g_s\gg 1$, which would
be consistent with the idea that the D-brane couplings to the NS5-brane
charge arise as a strong coupling effect.

If would also be interesting to investigate whether one can find
funnel solutions for fundamental strings, giving a dual description of 
the D$p{\perp}$F1 intersections. A possible framework for such an
analysis is provided by matrix strings \cite{dvv,schiappa}.

Another important aspect of the geometry in our constructions was
the spherical symmetry. In Section~4, this symmetry appeared 
essentially as an
assumption which we imposed in order to make the problem tractable.
Solving for symmetric instanton configurations
yields the curious relation
between $N$, the number of D-strings and $n$, the number of D5-branes:
$N\simeq n^3/6$ for large $N$. Remarkably, precisely the same relation arises
(again for large $N$) from the noncommutative geometry in the construction of
the D-string funnel --- see eq.~\reef{Nnrel}. Implicitly spherical
symmetry is again an underlying assumption in the latter construction,
\ie by definition the fuzzy four-sphere is $SO(5)$ symmetric \cite{wati4}.

We would expect that for any $N,n\gg1$ that the D5$\perp$D1 intersection 
has a smooth geometric resolution. However, it seems that the intricacies
of the non-abelian world volume theories will not allow for spherically
symmetric solutions in general. This is similar to the case of 
BPS $SU(2)$ magnetic monopoles, which describe
D-strings extending between parallel D3-branes at finite separation \cite{aki}.
It is known that there are no spherically symmetric solutions for
monopole charge greater than one. The best one can do is to obtain axially 
symmetric configurations \cite{manton}. On the D5-brane side, relaxing the
condition of spherical symmetry would enable one to obtain higher instanton
number solutions, but, of course, allowing for angular variations in the scalar
and gauge fields makes finding solutions significantly harder. Similarly,
on the D1-brane side relaxing the $SO(5)$ symmetry presumably allows for the
use of more  general matrices, but it is not at all clear how one should
proceed if the matrices $G^i$ are abandoned. We leave these issues for future
work.

\noindent{\bf SUSY and stability:}

There is, of course, one major distinction between the  D5$\perp$D1 system
and the D3$\perp$D1 system. That is, while the latter is supersymmetric,
the former is not. This is seen easily in perturbative string
theory \cite{tasi,clifford}, where one finds that having six Neumann-Dirichlet
directions for the D1-D5 strings is incompatible with supersymmetric ground
state. Given the lack of supersymmetry, one might worry that the configurations
describing the D5$\perp$D1 intersection are unstable. 
However, the inequality in eq.~(\ref{Renergy}) shows that the semi-infinite
fuzzy funnel is a minimum energy configuration, given the boundary
conditions that have been imposed. A similar inequality \reef{D5enbound}
appeared
in the discussion of the D5-brane spike. Here again we comment that in both
of world volume theories, the configurations describing the D5$\perp$D1
intersection
are singular but the singularities have an interesting physical interpretation
in terms of D-brane geometries. Now once we admit such non-standard or singular
boundary conditions, there is no obvious way in which the system could decay
compatible with the physics of D-branes. In other words, 
the boundary conditions define a certain super-selection sector, and since we
have the semi-infinite funnel to be the lowest energy configuration within this
sector, we can expect stability, at least to perturbative deformations of
the system. 

We note that there does exist a supersymmetric
configuration of a D1-brane orthogonal to a D5-brane where appropriate
fluxes of the NS two-form field are introduced on the D5-brane
\cite{vijay}. In fact by tuning these fluxes
the D1- and D5-branes can remain supersymmetric while at any angle.
The supersymmetry in these configurations is maintained
by the fact that the supersymmetric variation 
of the D5-brane world volume gaugino is altered in the presence of the
constant background B-field in such a way as to be compatible
with the supersymmetry of the D1-brane. These results would apply
most directly to an infinite D1-brane which ``pierces'' an orthogonal
D5-brane. To go from this system to the D5$\perp$D1 intersection considered
here, one would have to split the D-string on the D5-brane and move one half off to
infinity along the D5-brane. However, it is not obvious whether or not
supersymmetry could be maintained in this operation, but it would be
an interesting project for future research. One final observation is that
the supersymmetric configuration would not be spherically symmetric as the
B-field fluxes distinguish the different directions on the D5-brane
world volume.

\noindent{\bf Energy:}

\begin{figure}[htb]\label{pic}
\centerline{\includegraphics[width=10cm]{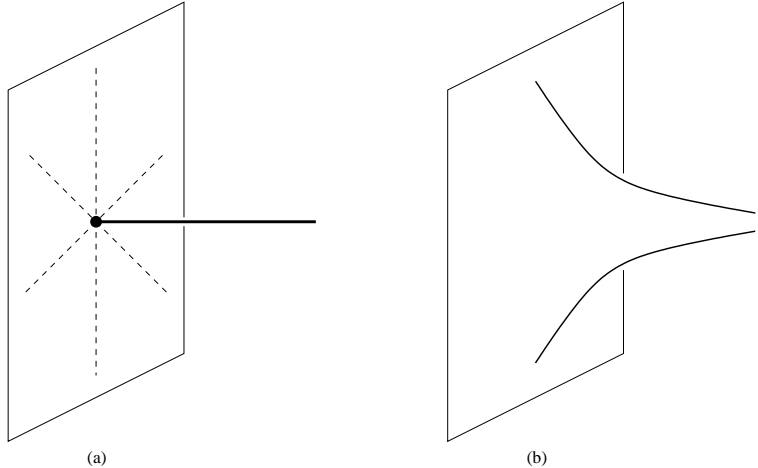}}
\caption{
(a) Schematic picture of strings ending on a brane. The endpoint of the
strings has a charge which is carried away by flux lines on the
brane. Alternatively one can think of the flux lines as a continuation
of the strings, smeared over the angular directions. 
(b) In reality the strings pull on the
brane, creating a spiked surface.}
\end{figure}

Being a non-supersymmetric state, the energy of the D5$\perp$D1 intersection
is not simply the sum of the energy of the constituents.
Perhaps surprisingly, the energy given in eqs.~\reef{energy} or \reef{D5en}
splits rather nicely into four distinct parts. The first and second
contributions correspond to the energy of the $n$ D5-branes and the $N$ semi-infinite
D-strings extending orthogonally away from the D5-branes. Next there is a
term that equals the energy of $N$ semi-infinite
D-strings extending out {\it radially} in the D5-brane world volume (like
the dashed lines in Fig.~\ref{pic} (a) below). Finally there is a 
finite binding energy, given in eq.~\reef{binde}.. 

How can we understand the appearance of the `radial' D-string term 
in the energy? Loosely speaking the D-strings do not terminate at
a fixed intersection point on the D5-brane, but rather they dissolve
into the five-brane world volume and spread out to infinity in the
guise of the gauge fields (\ref{chargeeq}). One can easily confirm that
the gauge configuration introduces the appropriate source for the RR two-form 
to correspond to the D-strings spreading out radially 
across the world volume. Of course, these comments would hold equally well
in discussing the bion spike describing D3$\perp$D1 intersection in Section~2.
Why then is this not reflected in the bion energy \reef{d3d1energy}?
If one considers parallel D-strings and D3-branes, a true bound state
is formed \cite{tasi,clifford,moreg}
but in the limit of large D3-brane volume, the energy
approaches that of the D3-branes alone. (The D-string charge is still
given by integrating the vanishingly small magnetic flux across the
entire D3-brane.) In contrast, parallel D-strings and D5-branes form
a marginal bound state whose total energy is always given by the sum of
that for each of the constituents. Hence, there remains a contribution 
in the energy from the radial spreading of D-strings. Of course, the
`radial' D-strings are not all parallel to each other, and so the
analogy to the supersymmetric system of parallel D-strings and D5-branes
is not precise. 

In fact one might expect a large additional contribution to the energy
due to the interaction between these `radial' D-strings. However, the
remaining term in the energy is a finite, negative contribution, $-\Delta
E$. This is the only term which distinguishes between
configurations in which the D-strings couple to the D5-brane 
as a single bundle or
not. For instance, consider a situation where we have $N$ D-strings
and a large number of D5-branes available for the D-strings to end
on. One possibility is to split the D-strings into two bundles with
$N_1$ and $N_2$ strings, each coupling to $n_1$ and $n_2$ D5-branes,
where $n_i\sim  N_i^{1/3}$ assuming large $n_i$ (using the
irreducible representation in the construction of the $SU(n_i)$
instantons). Since $\Delta E\sim
N(N/n)^{1/4}$ (see below eq.~(\ref{D5en})), the total binding 
energy for this
configuration is
\beq
\Delta E \sim \sum_i N_i^{7/6}.
\labell{binda}
\eeq
On the other hand, all of the D-strings can be combined in 
a single source of instanton number. This is accomplished
with roughly $N^{1/3}$~D5-branes, and the binding energy of such a 
configuration depends on $N$ as
\beq
\Delta E \sim (N^{7/6})=\left(\sum_i N_i\right)^{7/6}.
\labell{bindb}
\eeq
Hence this single bundle of D-strings has a larger binding energy than
the one 
in eq.~\reef{binda}. Noting that $\Delta E$ contributes with a negative sign 
to the total energy, we conclude that for $N$ D-strings and 
a large available number of D5-branes, we minimize the energy
(\ref{D5en}) by coupling all the D-strings to the smallest possible number $n$
of D5-branes compatible with the upper bound (\ref{suninst}).

It is also interesting to compare our results for the
D5$\perp$D1 intersection to the supergravity picture with
test D-strings in the background corresponding orthogonal D5-branes. 
The latter framework yields the correct energy for the D3$\perp$D1
system. The relevant fields for the D5-brane background are
given in eq.~(\ref{D5backg}). The energy of $N$ test D-strings
extending along $\sigma$ in this geometry is given by
\beq\labell{sugen}
E=N T_1\int_0^\infty d\sigma\,e^{-\phi}\sqrt{G_{tt}G_{\sigma\sigma}}
=N T_1\int_0^\infty d\sigma\sqrt{1+L^2/\sigma^2}.
\eeq
Here, of course, we find a divergent contribution as $\sigma\rightarrow\infty$
corresponding to the energy of the D-strings extending off to infinity.
Note, however, that there is a logarithmic divergence in the energy down
the throat of the D5-brane as $\sigma\rightarrow0$. 
This divergence emerges because of the nontrivial dilaton profile
in the D5-brane background (\ref{D5backg}). It is tempting to
associate this divergent energy with the contribution which we
identified with the radial D-strings above, but the precise form of the
divergences does not seem to agree. Even the leading order deviations from
flat space at large $\sigma$ disagree, similar what was found for the
fluctuations. We should also note that due to the variation of the
background dilaton, the effective string coupling is becoming smaller as
we approach the
D5-branes, and thus the effective D-string tension (\ie energy density)
is becoming large. At some point, the test brane picture of the D-strings
will break down, and there will be significant corrections to the
geometry. This suggests that the D5-brane supergravity background
is not the relevant one for this system. In fact, the supergravity
background (\ref{bakk}) we introduced in the analysis of the fluctuation
equations, also yields the correct test-string energy. This is to
be expected since the relations in eq.~(\ref{bakk}) ensured that the
quadratic actions agreed including the constant term. The latter essentially
corresponds to the energy density. It would be interesting to address
these issues in the context of the AdS/CFT along the lines of
ref.~\cite{lotsb}.

\noindent{\bf Other fuzzy funnels:}

In this paper, we have constructed the simplest extension of a new class of
brane intersections in analogy with the D3$\perp$D1 system. The smooth
resolution of the D5$\perp$D1 intersection owes its existence to the
non-abelian nature of the world volume gauge theories. It also provides
another example of the duality by which a given brane configuration can
be described by two different world volume theories. There are many
other extensions of this construction that might be considered.

For instance, one might construct fuzzy funnels with other noncommutative
geometries. For example, a number of different fuzzy geometries
are considered in ref.~\cite{coset}, including the coset geometries:
$SU(2)/U(1)\times SU(2)/U(1)$, $SU(3)/U(2)$ and $SU(3)/U(1)\times U(1)$.
While there the analysis focussed on the dielectric effect, 
their constructions can easily be adapted to produce new non-abelian
brane intersections. The first case above would simply yield a fuzzy
funnel with cross-section $S^2\times S^2$. Of course, the
interpretation would be that of a D1-brane intersecting a pair of 
mutually orthogonal D3-branes. The case of $SU(3)/U(2) \sim  CP^2$  would
give a four-dimensional cross-section and one can easily verify that
the funnel couples to the RR six-form. Hence this would also yield a
D5$\perp$D1 intersection but with a more exotic five-brane geometry.
Finally, the coset SU(3)/U(1)$\times$U(1) is six-dimensional and 
from the RR couplings, it appears the the corresponding funnel would describe
the D-strings expanding into a bound state of D7- and D5-branes.

A less exotic intersection which would be interesting is the
D7$\perp$D1 system. In this case, the system would be supersymmetric
and so perhaps the world volume field theory configurations would be
more reliable. In the D7-brane theory, one would have to consider
gauge field configurations on the six-sphere with nonvanishing third
Chern character. For the D-string description, the funnel would require
a fuzzy $S^6$. It may be that these configurations are related to the
work of ref.~\cite{selfd}.

Another interesting extension of our work would be to consider D-strings
joining collections of parallel D5-branes with finite separation. These
configurations would be related to the double funnels mentioned in 
Sections~3 and 4. However, given the results for the analogous configurations
of D-strings and D3-branes \cite{aki,bioncore}, it is likely that the true
minimum energy configurations will not be spherically symmetric. It would
also be interesting to consider the asymmetric case where the numbers of
D5-branes at either end of the D-string are not equal. This would lead
to exotic funnels similar to the domain walls considered
in ref.~\cite{pioline}.

Perhaps the ideas considered here could also be extended to  
even more exotic intersections in M-theory, involving M2-branes, M5-branes
and KK monopoles, using the non-abelian couplings proposed 
in ref.~\cite{lozano}.

\section*{Acknowledgments}
This research was supported by NSERC of Canada and Fonds FCAR du
Qu\'ebec.  RCM was also partially supported by the CIAR Cosmology and
Gravity Program.  We would especially like to thank Kiril Krasnov and
Jan Segert for helpful discussions. We would also like to acknowledge
interesting conversations with Bobby Acharya, Vijay Balasubramanian,
Marco Gualtieri, Nigel Hitchin, Neil Lambert, Washington Taylor and
Mark Van Raamsdonk. We also thank the Institute for Theoretical
Physics at UCSB for its hospitality during the early stages of this
work.

\appendix
\section{The $G^i$ matrices}

For the convenience of the reader, we here reproduce several formulae
involving the matrices $G^i$ which were used to construct our D-string
funnel. These matrices were presented in ref.~\cite{wati4}
to provide a fuzzy four-sphere. They satisfy certain
physical properties including producing a spherical locus,
$SO(5)$ rotational invariance and an
appropriate spectrum of eigenvalues.
The equations below are taken directly from the paper
by Castelino, Lee and Taylor \cite{wati4}, and we refer the
reader there for more details. 

The matrices $G^i$ and $G^{ij}$ are defined as follows:
\begin{eqnarray}
G^i &=& \left(
\Gamma^i \otimes \identity \otimes \cdots \otimes \identity +
\identity \otimes \Gamma^i \otimes \identity \otimes 
\cdots \otimes \identity + \cdots +
\identity\otimes \cdots \otimes\identity \otimes\Gamma^i \right)_{\rm Sym},
\cr
G^{ij} &=& {1\over2} [G^i,G^j]  ,
\end{eqnarray}
where $\Gamma^i$, $i=1,\ldots,5$ are $4 \times 4$ Euclidean gamma
matrices,
and $\identity$ is the identity matrix. The subscript Sym means
the matrices are restricted to the completely symmetric tensor product space.
An explicit construction of
these matrices in $2 \times 2$ block-diagonal form is given by 
\begin{eqnarray}
\Gamma^j & = &  \left(\begin{array}{cc}
0 & -i \sigma_j\\
i \sigma_j& 0
\end{array}\right), \;\;\;\;\; j \in\{1, 2, 3\}, \nonumber\\
\Gamma^4 & = &  \left(\begin{array}{cc}
0 & \identity_2 \\
\identity_2& 0
\end{array}\right),\\
\Gamma^5 & = &  \left(\begin{array}{cc}
\identity_2 & 0\\
0&  -\identity_2
\end{array}\right), \nonumber
\end{eqnarray}
where $\sigma_j$ are the usual Pauli matrices.

The dimension and the ``Casimir'' of the $n$-fold totally symmetric 
representation $G^i$ are
\beq
N=\frac{(n+1)(n+2)(n+3)}{6},\qquad  \ c = n (n+4).
\eeq
The Casimir, $c$, appears in the following identities:
\begin{equation}
G^i G^i = c\, \identity_N, \;\;\;\;\;\  \ G^{ij} G^{ji} = 4 c\, \identity_N.
\labell{gicas}
\end{equation}
The commutators of $G^i$'s and $G^{jk}$'s are easily obtained from those of 
gamma matrices:
\begin{eqnarray}
[G^{ij},G^k] &=& 2(\delta^{jk} G^i - \delta^{ik} G^j),
\labell{comme}\cr
[G^{ij}, G^{kl}] &=& 
2(\delta^{jk}G^{il} + \delta^{il}G^{jk} 
- \delta^{ik}G^{jl} - \delta^{jl}G^{ik}).
\end{eqnarray}

Note then that the $G^{ik}$ are the generators of $SO(5)$ rotations.
There is a very useful identity between the anti-commutators: 
\beq
\{ G^{ij}, G^{jk} \} + \{ G^i, G^k \} = 2c\,\delta^{ik} \identity_N,
\labell{idum}
\eeq
and we also need the following identities:
\beqa
G^{ij}G^j&=&4G^i, 
\nonumber \\
G^{ij}G^{jk}&=&c\delta^{ik}+G^iG^k-2G^kG^i,
\nonumber \\
\epsilon^{ijklm}  G^i G^j G^k G^l &=& (8n + 16) G^m.
\labell{GGGGeq}
\eeqa

Combining the above definitions and identities, it is straightforward to derive
the following formulae
\begin{eqnarray*}
G^i G^j G^i G^j = (c^2 - 8c)\identity_N,  &\hspace{0.5in} &
G^i G^j G^{ik} G^{kj} = -8c\,\identity_N,
\\
G^i G^{jk} G^i G^{jk} =(-4 c^2 + 16 c)\identity_N,  & &
G^{ij} G^{kl} G^{ij} G^{kl} =(16c^2 - 96 c)\identity_N,
\\
G^{ij} G^{jk} G^{kl} G^{li} = (4c^2 + 32c)\identity_N, & &
G^{ij} G^{jk} G^{il} G^{lk} = (4c^2 - 40c)\identity_N,
\end{eqnarray*}
\[
G^iG^jG^{jk}G^{ki}=G^iG^{ik}G^jG^{jk}=G^iG^{jk}G^jG^{ik}=16c\identity_N.
\]

\section{Homogeneous instantons\protect\footnotemark}
\footnotetext{We would like to acknowledge
Jan Segert for invaluable assistance on the material presented
in the following.}

It is easy to construct self-dual $SU(2)$ gauge fields which are
homogeneous on $S^4$ with the usual round metric and which have instanton
number $N=1$. The construction takes the instanton on $\bbR^4$ with unit
scale size and makes a stereographic projection to $S^4$. The
gauge field of an $SU(2)$ instanton located at the origin in $\bbR^4$, 
with scale size $\mu$, is given by \cite{bpst}
\beq
iA_\mu(x)={|x|\over\mu^2+|x|^2}\left(x^4+ix^a\sigma_a\right)
\partial_\mu\left({x^4-ix^a\sigma_a\over |x|}\right),
\eeq
where $\sigma_a$ ($a=1,2,3$) denotes the Pauli matrices, interpreted
as $SU(2)$ generators acting on the fundamental representation. The explicit
components then read
\beq
A_a(x)={\epsilon_{abc}x^c\sigma_b-x^4\sigma_a\over\mu^2+|x|^2},\quad
A_4(x)={x^a\sigma_a\over\mu^2+|x|^2}
\ ,\labell{gaugf}
\eeq
while the corresponding field strength is
\beq
F_{ab}={2\mu^2\epsilon_{abc}\sigma_c\over(\mu^2+|x|^2)^2},\ \ 
F_{a4}={2\mu^2\sigma_a\over(\mu^2+|x|^2)^2}\ ,
\eeq
which is explicitly self-dual.

The stereographic projection from $\bbR^4$ to $S^4$ is accomplished by
the map
\beqa
x^1&=&\cot(\alpha^1/2) c_2,\nonumber\\
x^2&=&\cot(\alpha^1/2) s_2 c_3,\nonumber\\
x^3&=&\cot(\alpha^1/2) s_2 s_3 c_4,\nonumber\\
x^4&=&\cot(\alpha^1/2) s_2 s_3 s_4,\nonumber
\eeqa
where $s_i\equiv\sin(\alpha^i)$, $c_i\equiv\cos(\alpha^i)$ and
the angles on the four-sphere are defined in eq.~\reef{jdsuni}.
Performing this coordinate
transformation for the $\mu=1$ instanton, we find the explicit field
strengths
\beqa
\tF_{12}&=&{\lambda\over2}s_1[s_3s_4\sigma_1-s_3c_4\sigma_2+c_3\sigma_3],
\nonumber\\
\tF_{13}&=&{\lambda\over2}s_1s_2[(s_2c_4+c_2c_3s_4)\sigma_1
+(s_2s_4-c_2c_3c_4)\sigma_2-c_2s_3\sigma_3],\labell{hominst}\\
\tF_{14}&=&{\lambda\over2}s_1s_2s_3[(c_2c_4-s_2c_3s_4)\sigma_1
+(c_2s_4+s_2c_3c_4)\sigma_2+s_2s_3\sigma_3],\nonumber
\eeqa
where as in Section~4, we use the definition $\tF_{ij}\equiv\lambda
F_{\alpha^i\alpha^j}$. The remaining components of the field strength
follow from self-duality, {\it e.g.},
$\tF_{34}=\sqrt{g}g^{11}g^{22}\tF_{12}=s_1s_2^2s_3\tF_{12}$.
This configuration manifestly satisfies the condition for a homogeneous
instanton (\ref{fdfeq}) with $N=1$,
\beq
\Tr \tF_{ij}\dtF^{ij}=6\lambda^2 .
\eeq
In fact, we note that as required at various steps in the analysis of
the D5-brane spike, this expression is proportional to the gauge
identity matrix, \ie $\tF_{ij}\dtF^{ij}=3\lambda^2\,\identity_2$.

The above gauge fields can be extended to a homogeneous instanton 
for the gauge group $SU(n)$ by simply replacing the $\sigma_a$ in
eqs.~(\ref{gaugf}) and (\ref{hominst}) by an $n\times n$ matrix representation
of the $SU(2)$ algebra (\ref{su2alg}). The latter can then be interpreted
as $SU(n)$ generators acting in the fundamental representation.
If we choose the irreducible $n\times n$ representation,
the corresponding gauge field satisfies
\beq
\tF_{ij}\dtF^{ij}=(n^2-1)\lambda^2\,\identity_n,
\eeq
corresponding to instanton number (and D-string number)
\beq
N={n(n^2-1)\over6}.
\eeq
A further result is that the homogeneous $SU(n)$ gauge field with
the maximum possible instanton number is in fact the above configuration
constructed with the irreducible $n\times n$ representation of $SU(2)$.
The proof, which we sketch below,
relies on two mathematical results. The first is
is the classification of homogeneous principal bundles, and the
second is the classification of symmetric connections on the
homogeneous principal bundles, by a theorem of H.C.~Wang \cite{wang}. 

In general, we wish to consider
principal fibre bundles which admit a fibre-transitive Lie group of
automorphisms, which will be a necessary ingredient for Wang's
theorem. Let $P$ be a principal $H$ bundle and $M=P/H$ is the base
space. Let $G$ be the automorphism group which acts on $P$ by
left multiplication. These transformations commute with the $H$
action (on the right), and so induce a $G$ action on $M$. Suppose that
this $G$ action on $M$ is transitive, and that $G_0$ is the stabilizer
at some point $x$ in $M$. Then $G_0$ acts on the fibre over $x$, commuting
with the $H$ action. Since the fibre of the principal bundle is isomorphic
to the group $H$, the $G_0$ action must be via a homomorphism
$\phi:G_0\rightarrow H$. Thus the homogeneous principal bundles
are in one-to-one correspondence with the latter homomorphisms.
For the particular case at hand, we consider $H=SU(2)$, $M=S^4$
and $G=Spin(5)=Sp(4)$. Now the stabilizer is $G_0=Spin(4)\simeq
SU(2)\times SU(2)$, and hence we consider homomorphisms
$\phi:SU(2)\times SU(2)\to SU(2)$. Aside from the trivial case, there
are  two homomorphisms: one which is the identity map on the first factor and
trivial on the second, and another which switches the role of the two
$SU(2)$ factors. Only one of these supports a self-dual connection
(the other is anti-self-dual), and it corresponds to the quaternionic
Hopf fibration of the seven-sphere \cite{ssss}, \ie $S^7$ as $SU(2)=S^3$
fibred over $S^4$. For a principal $SU(n)$ bundle, we have the
homomorphisms $\phi:SU(2)\times SU(2)\to SU(n)$. Now these homomorphisms
will split as $\phi_1:SU(2)\to K_1\subset SU(n)$ and
$\phi_2:SU(2)\to K_2\subset SU(n)$. Hence all of the
homogeneous ($Spin(5)$ invariant) bundles will decompose as bundles
associated with the Hopf and anti-Hopf fibrations. Now given a
$G$-invariant principal $H$ bundle, Wang's theorem \cite{wang}
puts the invariant connections over these bundles in one-to-one
correspondence with certain linear mappings from the Lie algebras
of $G\to H$. Essentially these are the mappings consistent with
the homomorphisms identified above. Hence in the case of interest,
identifying the invariant connections corresponds to identifying
$SU(2)$ subgroups of $SU(n)$. That is, the above construction
in fact includes all of the homogeneous gauge configurations
on the four-sphere. It is straightforward to show then that the
instanton number is maximized by the choice of the irreducible
$n\times n$ representation.

\end{document}